\documentclass[twocolumn,showpacs,aps,prl,superscriptaddress]{revtex4}

\usepackage{graphicx}
\usepackage{dcolumn}
\usepackage{amsmath}
\usepackage{multirow}

\input{./pubboard/babarsym}

\def\fish    {\ensuremath{\cal F}\xspace}
\def\fishlt    {\ensuremath{\cal F_{T}}\xspace}

\def\qq {\ensuremath{q\bar{q}}\xspace}
\def\alphaeff {\ensuremath{\alpha_{\rm eff}}\xspace}
\def\hpm    {\ensuremath{h^{\pm}}\xspace} 

\def\Btopipi   {\ensuremath{\B \to \pi\pi}\xspace}
\def\Btopipiz   {\ensuremath{\Bpm \to \pipm\piz}\xspace}
\def\Btokpiz   {\ensuremath{\Bpm \to \Kpm\piz}\xspace}
\def\Btokzpi   {\ensuremath{\Bpm \to \Kz\pipm}\xspace}
\def\Btohpiz   {\ensuremath{\Bpm \to \hpm\piz}\xspace}
\def\Btorhopiz   {\ensuremath{\Bpm \to \rho^{\pm}\piz}\xspace}
\def\Bztopizpiz   {\ensuremath{\Bz \to \piz\piz}\xspace}

\def\Btorhopi   {\ensuremath{\B \to \rho\pi}\xspace}

\def\de {\ensuremath{\Delta E}\xspace}
\def\cossph   {\ensuremath{|\cos{\theta_{\scriptscriptstyle S}}|\;}\xspace}

\def\acp {\ensuremath{{\cal A}}\xspace}
\def\acpcp {\ensuremath{{\cal A}_{\CP}}\xspace}

\def\mz        {\mbox{$m_{0}$}\xspace}
\newlength{\LL}
\newcommand{\etal}{{\it et al.}}

\long\def\inst#1{\par\nobreak\kern 4pt\nobreak
    {\it #1}\par\vskip 10pt plus 3pt minus 3pt}

\begin{document}

\preprint{\babar-PUB-03/002}
\preprint{SLAC-PUB-9683}

\begin{flushleft}
\babar-PUB-03/002 \\
SLAC-PUB-9683\\
\end{flushleft}

\title{
{\Large \bf \boldmath
Observation of the Decay \Btopipiz , Study of \Btokpiz , and Search for \Bztopizpiz
}
}

%
\author{B.~Aubert}
\author{R.~Barate}
\author{D.~Boutigny}
\author{J.-M.~Gaillard}
\author{A.~Hicheur}
\author{Y.~Karyotakis}
\author{J.~P.~Lees}
\author{P.~Robbe}
\author{V.~Tisserand}
\author{A.~Zghiche}
\affiliation{Laboratoire de Physique des Particules, F-74941 Annecy-le-Vieux, France }
% ...
\collaboration{The \babar\ Collaboration}
\noaffiliation

\date{\today}

\begin{abstract}
  
  We present results for the branching fractions and charge
  asymmetries in \Btohpiz (where $\hpm$ = $\pipm, \Kpm$) and a search
  for the decay \Bztopizpiz using a sample of approximately $88$
  million \BB\ pairs collected by the \babar\ detector at the \pep2\ 
  asymmetric-energy $B$ Factory at SLAC.  We measure ${\BR}
  (\Btopipiz) = (5.5 ^{+1.0}_{-0.9} \pm 0.6) \times 10^{-6}$, where
  the first error is statistical and the second is systematic.  The
  \Btopipiz signal has a significance of $7.7 \sigma$ including
  systematic uncertainties.  We simultaneously measure the $\Kpm\piz$
  branching fraction to be ${\BR}(\Btokpiz) = (12.8 ^{+1.2}_{-1.1} \pm
  1.0) \times 10^{-6}$. The charge asymmetries are $\acp_{\pipm\piz} =
  -0.03 _{-0.17}^{+0.18} \pm 0.02$ and $\acp_{\Kpm\piz} = -0.09 \pm
  0.09 \pm 0.01$.  We place a 90\% confidence-level upper limit on the
  branching fraction ${\BR}(\Bztopizpiz)$ of $3.6 \times 10^{-6}\,$.

\end{abstract}


\maketitle

The study of \B meson decays into charmless hadronic final states
plays an important role in the understanding of \CP violation in the
\B system. In the Standard Model, \CP violation arises from a single
complex phase in the Cabibbo-Kobayashi-Maskawa quark-mixing matrix $V_{\rm ij}$~\cite{ckmref}. 
Measurements of the time-dependent \CP-violating asymmetry in the
\Bztopipi decay mode by the \babar~\cite{babarsin2alpha}
and Belle~\cite{bellesin2alpha} collaborations  provide information on the angle 
$\alpha \equiv \arg\left[-V_{\rm td}^{}V_{\rm tb}^{*}/V_{\rm ud}^{}V_{\rm ub}^{*}\right]$ of
the Unitarity Triangle. However, in contrast to the theoretically
clean determination of the angle $\beta$ in \Bz decays to charmonium
final states~\cite{babarsin2beta,bellesin2beta}, the extraction of
$\alpha$ in \Bztopipi is complicated by the interference of tree and
penguin amplitudes with different weak phases.  The shift between \alphaeff,
from the measured \Bztopipi asymmetry, and
$\alpha$ may be evaluated or constrained using measurements
of the isospin-related decays $\Bz (\Bzb) \to \piz\piz$ and
$\Bpm\to\pipm\piz$~\cite{Isospin}.

The \CP-violating charge asymmetry for \Bpm modes, defined as 
\begin{equation}
 \acpcp \equiv \frac{|\bar{A}|^2 - |A|^2}{|\bar{A}|^2 + |A|^2},
\end{equation}
where $A$ ($\bar{A}$) is the \Bp (\Bm) decay amplitude, will deviate
from zero if the tree
and penguin amplitudes each have different weak and strong phases. In
the Standard Model  the
decay \Btopipiz has only a tree amplitude contribution, so no charge
asymmetry is expected.  
Both the rate and asymmetry of the decay \Btokpiz may constrain the
value of the Unitarity Triangle angle $\gamma$. In particular, the ratio of ${\BR}(\Btokpiz)$ and
${\BR}(\Btokzpi)$ provides a lower bound for $\gamma$~\cite{GronauRosner}.
The decay \Btokpiz can also exhibit a significant charge
asymmetry; different models for hadronic \B decays predict a range of
values~\cite{Asymmetry}.

 In this
paper, we report on an observation of the decays  \Btopipiz and \Btokpiz,
a measurement of their \CP-violating charge asymmetries, and a search for
the decay \Bztopizpiz, using $(87.9\pm 1.0) \times 10^{6}$ \BB pairs collected
with the \babar\ detector. 

\babar\ is a solenoidal detector optimized for the asymmetric-energy
beams at \pep2 and is described in detail in Ref.~\cite{babarnim}.
Charged particle (track) momenta  are measured with a
5-layer double-sided silicon vertex tracker (SVT) and a 40-layer drift
chamber (DCH) inside  a 1.5 T superconducting solenoidal magnet.
Photon (neutral cluster) positions and energies are measured with an
electromagnetic calorimeter (EMC) consisting of 6580 CsI(Tl) crystals.
Tracks are identified as pions or kaons by the Cherenkov angle
$\theta_c$ measured with a detector of internally reflected Cherenkov 
light (DIRC).


High efficiency for recording \BB events in which one \B decays with low 
multiplicity  is achieved with a two level trigger with complementary
tracking and calorimetry-based trigger decisions. 
\BB events are selected using track and neutral cluster content and
event topology.  

\begin{table*}[!htb]
\begin{center}
\caption{The results for both \Btohpiz and \Bztopizpiz are summarized.  The number
of \B candidates $N$, total detection efficiencies $\epsilon$,  fitted signal 
yields $N_{S}$, significances $S$,  
charge-averaged branching fractions \BR, asymmetries ${\cal A}$, and $90\%$ C.L. asymmetry limits are shown.
Errors are statistical and systematic respectively, with the exception of $\epsilon$ whose 
error is purely systematic.
The upper limit for the \Bztopizpiz branching fraction corresponds to the $90\%$ C.L., and
the central value is shown in parentheses.}
\label{table:summary}
\begin{ruledtabular}
\begin{tabular}{lccccccc} 
Mode  & $N$ & $\epsilon$ (\%) & $N_{S}$ & $S(\sigma)$ & \BR($10^{-6}$) & ${\cal A}$ & ${\cal A}(90\% {\rm C.L.})$  \\ 
\hline \\[-3mm]
$\pipm\piz$ & \multirow{2}{\LL}{21752} & $26.1 \pm 1.7$ & 
              $ 125_{-21}^{+23}  \pm 10$      & 7.7  & $ 5.5 _{-0.9}^{+1.0} \pm 0.6  $ & $-0.03 _{-0.17}^{+0.18} \pm 0.02$ & $[-0.32,0.27]$ \\[1mm]
$\Kpm\piz$  &                           & $28.0 \pm 2.0$ & 
              $ 239 ^{+21}_{-22} \pm 6$       &  17.4    & $ 12.8 _{-1.1}^{+1.2} \pm 1.0 $ & $-0.09 \pm 0.09 \pm 0.01 $        & $[-0.24,0.06]$  \\[1mm]
$\piz\piz$  & 3020                      & $16.5 \pm 1.7$ & 
              $ 23 ^{+10}_{-9} \,^{+8}_{-4} $ & 2.5  &  $<3.6 \, (1.6^{+0.7}_{-0.6} \,^{+0.6}_{-0.3})$ &  \\
\end{tabular}
\end{ruledtabular}
\end{center}
\end{table*}

Candidate \piz\ mesons are reconstructed as pairs of photons, spatially
separated in the EMC, 
with an invariant mass within $3 \sigma$ of the  \piz\ mass.
The resolution sigma is approximately 8~\mevcc for high momentum \piz.  
Photon candidates are required to be consistent with
the expected lateral shower shape, not be matched to a track, and have a
minimum energy of 30 \mev. 
To reduce the background from false \piz candidates, the 
angle $\theta_{\gamma}$ between the photon momentum vector in the \piz rest frame and
the \piz momentum vector in the laboratory frame is required to satisfy
$|\cos{\theta_{\gamma}}| < 0.95$.  The \piz\ candidates are 
fitted kinematically with their mass constrained to the nominal \piz 
mass.  

Candidate tracks are required to be within the
tracking fiducial volume, originate from the interaction point, 
consist of at least 12 DCH hits, and be associated with at least 6 Cherenkov
photons in the DIRC. 

\B meson candidates are reconstructed by combining a 
\piz with a pion or kaon (\hpm) or by combining two \piz mesons. 
Backgrounds arise from two sources: \Btorhopi decays in which one pion is emitted nearly at rest in the \B
frame so that the remaining decay products are kinematically consistent
with a \Btopipiz or \Bztopizpiz  decay, and $\epem \to \qqbar \,(q = u,d,s,c)$ events where
an \hpm or \piz from each quark randomly combine to mimic a \B decay.

Both backgrounds are separated from signal 
using the kinematic constraints of \B mesons produced at the \FourS. The
first kinematic variable is the beam-energy substituted mass $\mes =
\sqrt{ (s/2 + {\bf  p}_{i}\cdot{\bf p}_{B})^{2}/E_{i}^{2}- {\bf p}^{2}_{B}}$,
where  $\sqrt{s}$ is the total center-of-mass (CM) energy.
$(E_{i},{\bf p}_{i})$ is the four-momentum of the initial \epem
system  and ${\bf p}_{B}$ is the \B momentum
both in the laboratory frame. The second variable is $\de = E_{B} -
\sqrt{s}/2$, where $E_{B}$ is the \B candidate energy in
the CM frame.  The pion mass is assigned to all \hpm candidates for the \de calculation. 

The \Btorhopiz
background to \Bztopizpiz is reduced by only using candidates
with  $|\de| < 0.2\gev$. Remaining \Btorhopiz background is
further suppressed by removing candidates in which the additional \pipm is identified.
The track that gives a $\pipm\piz$ invariant mass and
$\mes$ of the $\pipm\piz\piz$ combination most consistent with the
$\rho$ and \B mass is selected.  Requirements on the resulting
$\pipm\piz$ invariant mass and  on the \de of the $\pipm\piz\piz$
combination remove roughly 50\% of the remaining \Btorhopiz
background, with 93\% efficiency for \Bztopizpiz.
Only $(0.40 \pm 0.04)\%$ of \Btorhopiz decays, and a negligible
fraction of nonresonant $\Bpm \to \pipm\piz\piz$ decays,  remain  after all
cuts.
For \Btohpiz the \Btorhopi background is suppressed by
selecting candidates with   $ -0.11 < \de < 0.15 \gev $.

The jet-like \qqbar background is suppressed by
requiring that the angle $\theta_{\scriptscriptstyle S}$ between the sphericity axes of
the \B candidate and of the remaining tracks and
neutral clusters in the event, in the CM frame, satisfy $\cossph < 0.8 \,(0.7)$
for \Btohpiz (\Bztopizpiz).  Also, we require $\mes > 5.2 \gevcc$.
The number of \Btohpiz and \Bztopizpiz candidates satisfying these requirements and the
estimated efficiencies, obtained from simulated data, 
are shown in the first two columns of
Table~\ref{table:summary}. The simulation has been tuned to reproduce
the observed
track and \piz efficiencies. The error in the estimated efficiency
is dominated by the $5\%$ systematic uncertainty in the single \piz reconstruction efficiency.

The number of signal \B candidates is determined in an extended unbinned
maximum likelihood fit.  The probability ${\cal P}_i\left(\vec{x}_j;
  \vec{\alpha}_i\right)$ 
for a signal or background hypothesis
is the product of probability density functions (PDFs) for the
variables $\vec{x}_j$  given the set of parameters $\vec{\alpha}_i$.
The likelihood function is given by a product over all $N$ events and
the $M$ signal and background hypotheses:
\begin{equation}
{\cal L}= \exp\left(-\sum_{i=1}^M n_i\right)\,
\prod_{j=1}^N \left[\sum_{i=1}^M N_i {\cal P}_i\left(\vec{x}_j;
\vec{\alpha}_i\right)
\right]\, .
\end{equation}
For \Btohpiz the probability coefficients are  $N_{i} = \frac{1}{2}(1-
q_j \acp_i) n_i $, where  $q_j$ is the
charge of the track $h$ and  the fit parameters $n_i$ and $\acp_i$ are the number
of events and asymmetry for the  four $\pip\piz$ and $\Kp\piz$ signal and
background components.
For \Bztopizpiz the coefficients are $N_{i} = n_i$ where the three $n_i$ are the
number of signal candidates, \Btorhopiz background and \qqbar background.
Monte Carlo simulations are used to verify that the likelihood fits are unbiased.

The variables $\vec{x}_j$ used for \Btohpiz are \mes, \de, the
Cherenkov angle $\theta_c$ of the \hpm track, and a Fisher discriminant \fish.  The Fisher
discriminant is given by an optimized linear combination of $\sum_i
p_i$ and $\sum_i p_i |\cos{\theta_i}|^2$
where $p_{i}$ is the momentum and $\theta_{i}$ is the angle with
respect to the thrust axis of the \B candidate, both in the CM frame,
for all tracks and neutral clusters not used to reconstruct the \B
meson. 

The PDFs for \mes, \de, $\theta_c$, and \fish for the background are
determined using data, while the PDFs for signal are found from a
combination of simulated events and data.  The \mes distribution for
background is modeled as a threshold function~\cite{Argus}, whose
shape parameter is a free parameter of the fit. The \de distribution
for background is modeled as a quadratic function whose parameters are
determined from the \mes sideband in data. The \mes and \de
distributions for signal are modeled as Gaussian distributions with a low-side
power-law tail whose parameters are found with simulated
events.  The \de resolution is approximately 42~\mev based on
simulated events and is confirmed by evaluating the resolution in a
sample of $\Bpm \to \Dz \rho^{\pm} \, (\rho^{\pm} \to \pipm \piz)$
events with an energetic \piz.  To allow for EMC energy scale
variations, the mean of the \de PDF is a free parameter of the
fit.  To account for the use of the pion mass hypothesis, the mean of
\de is shifted for the $\Kpm\piz$ PDFs.  The \fish distribution is
modeled as a bifurcated Gaussian and a double Gaussian for signal and
background respectively, whose parameters are determined for signal
from simulation and for background from \mes sidebands. The difference of the measured and
expected values of $\theta_c$ for the pion or kaon hypothesis,
divided by the uncertainty on $\theta_c$, is modeled as a double
Gaussian function. 
A control sample of kaon and pion tracks, from the decay $\Dstarp \to
\Dz \pip \, , \Dz \to \Km \pip$, is used to parameterize
$\sigma_{\theta_c}$ as a function of the track polar angle.

The variables $\vec{x}_j$ used for \Bztopizpiz are \mes, \de, and
another Fisher discriminant \fishlt. The \fishlt combines \fish with
information from the \B tagging algorithm described in
Ref.~\cite{babarsin2beta}.  The tagging algorithm uniquely classifies
events according to their lepton, kaon, and slow pion (from $\Dstarp
\to \Dz \pip_{\rm slow}$) content, using all tracks in the event.
Nine event classes, in decreasing order of their
background rejection, contain the following: a high momentum electron
and a kaon, a high momentum muon and a kaon, a high momentum electron,
a high momentum muon, a kaon and a slow pion, a well identified kaon,
a slow pion, any kaon, or none of the above.  These event classes are
assigned an index, which is a new discriminating variable, and is
combined with \fish into a second Fisher discriminant \fishlt,
optimized using simulated events.

\begin{figure}[!tbp]
\begin{center}
  \includegraphics[width=0.49\linewidth]{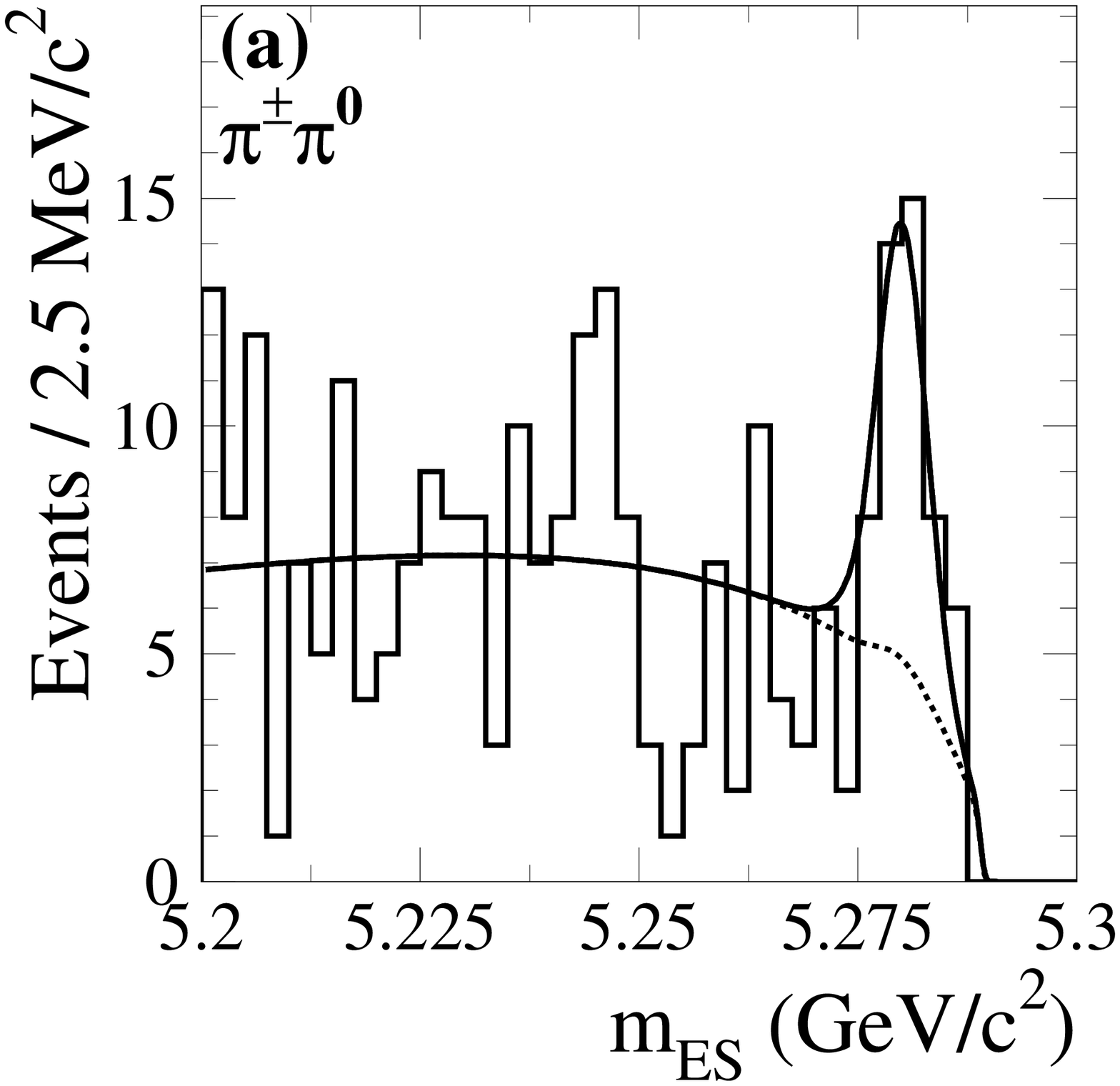}
 \includegraphics[width=0.49\linewidth]{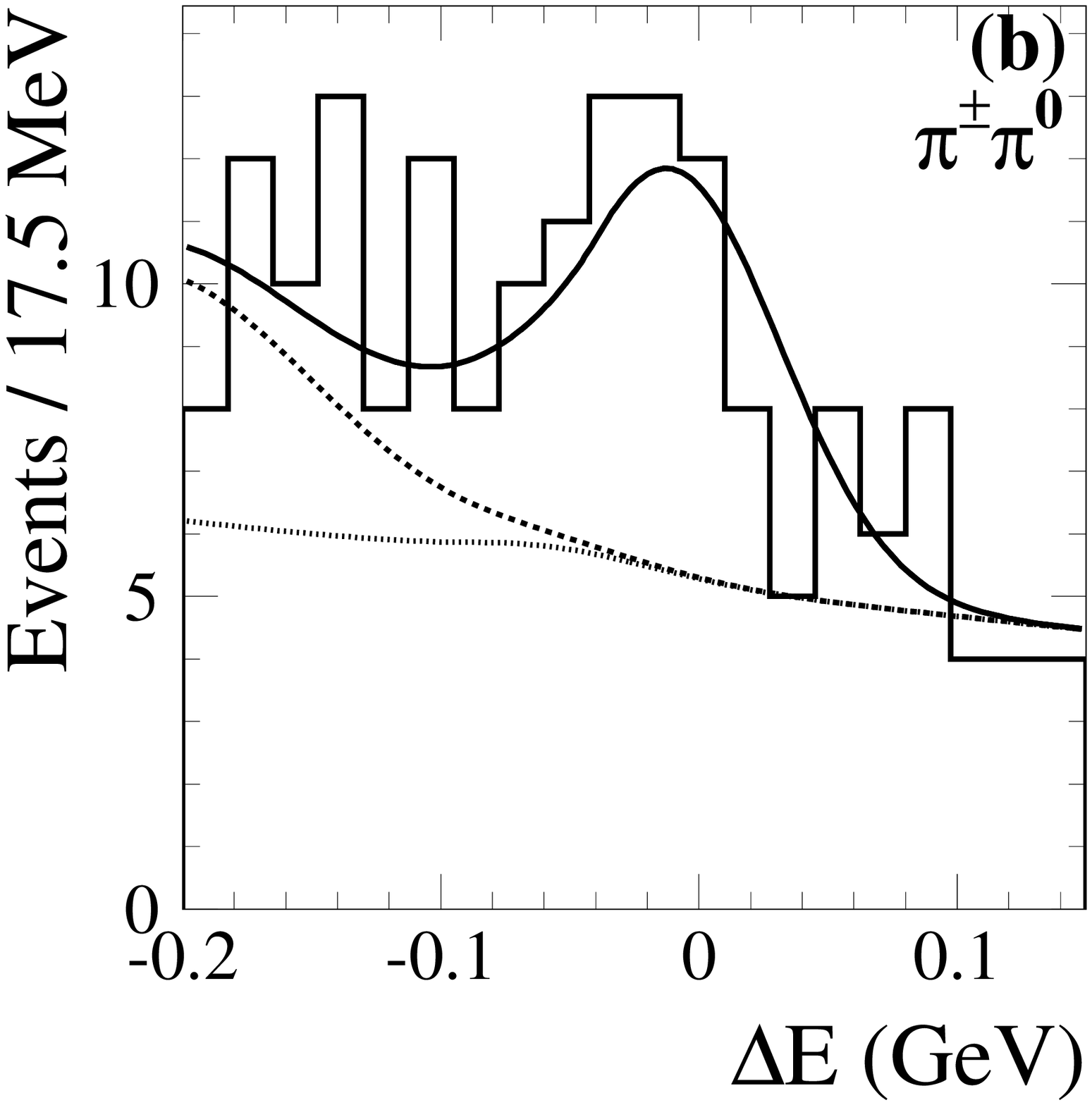}
\vspace{-0.9cm}
  \includegraphics[width=0.49\linewidth]{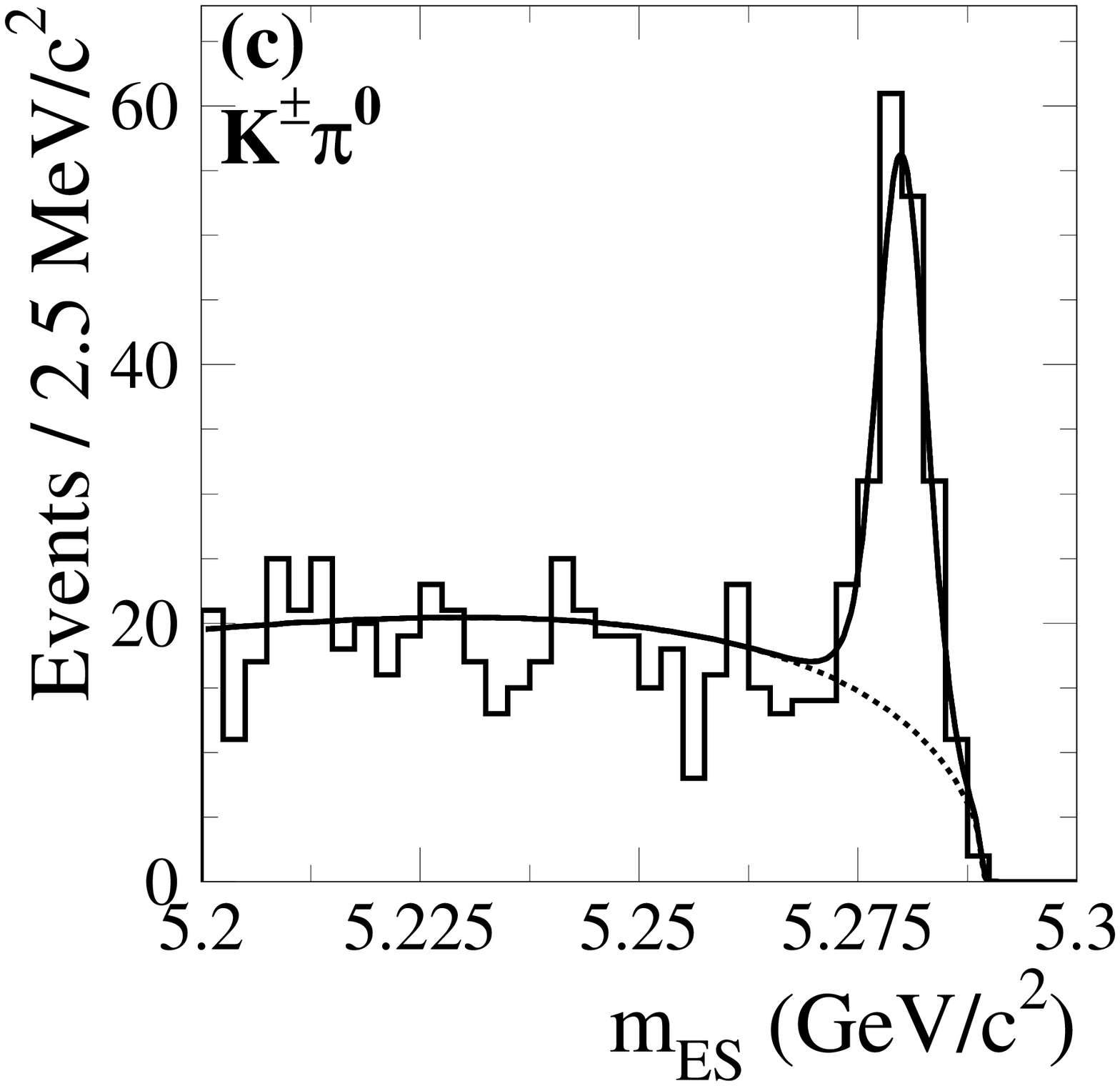}
  \includegraphics[width=0.49\linewidth]{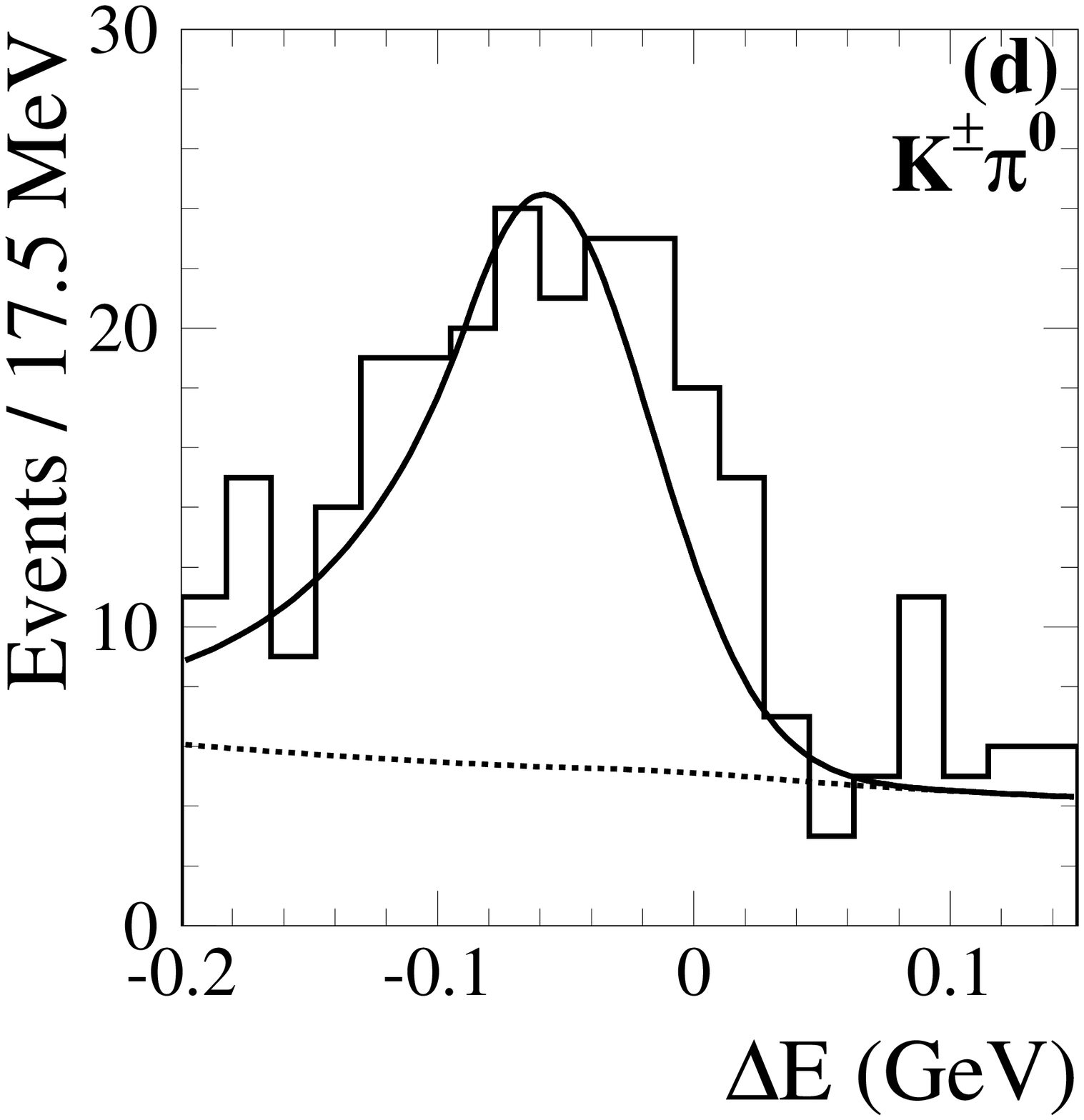}
\end{center}
\vspace{0.2cm}
\caption{The distributions of \mes (left) and \de (right) 
for \Btopipiz (top) and \Btokpiz  (bottom), for
candidates that satisfy optimized requirements on probability ratios
for signal to background  based on all variables except the one being
plotted. 
The fraction of signal events included in the plots is 24\% (\mes) and 35\% (\de)
for $\pipm\piz$,  and 53\% (\mes) and 48\% (\de) for $\Kpm\piz$.
Solid curves represent projections of the complete maximum 
likelihood fit result; dotted curves represent the background 
contribution.  For the \Btopipiz \de distribution, the dotted curve
shows the \qqbar background and the small \Btokpiz cross-feed;
the dashed curve includes the \Btorhopi background as well, so is the
sum of all backgrounds.}
\label{fig:hpiz}
\end{figure}

 The \mes distribution for
\qqbar background is parameterized by the same threshold
function used in the \Btohpiz analysis,
where the shape parameter is determined from data with
$\cossph > 0.9$.  The \de distribution for \qqbar background is modeled
as a
quadratic polynomial with parameters found from on-resonance data in the
\mes sidebands and off-resonance data.  The \mes and \de variables in
both \Bztopizpiz and \Btorhopiz are correlated, so a two dimensional
PDF derived from a smoothed simulated distribution is used.  The \de
resolution is approximately 80~\mev. The
\fishlt distribution for \qqbar, \Btorhopiz,  and \Bztopizpiz
is modeled as the sum of three Gaussians.
For \qqbar the parameters are found using both \mes sideband and off-resonance data.
For \Bztopizpiz and \Btorhopiz the parameters are found using 
a sample of fully reconstructed $\Bz \to D^{(*)} n\pi \,(n=1,2,3) $
events.

The decay \Btorhopiz has not been observed; Ref.~\cite{cleorhopi}  set
an upper limit of $\mathcal{B}(\Btorhopiz) < 4.3\times10^{-5}$ at 90\%
C.L. based on a measured central value of $\mathcal{B}(\Btorhopiz) = 2.4\times10^{-5}$.  
Therefore we fix the number of \Btorhopiz events in the fit to 
$n_{\rho\piz} = 8.4$, based on this central value, and evaluate the systematic uncertainty of
allowing  $n_{\rho\piz}$ to vary from 4.2 to 15 events.

\begin{figure}[!tbph]
\begin{center}
\includegraphics[width=0.455\linewidth,height=0.482\linewidth]{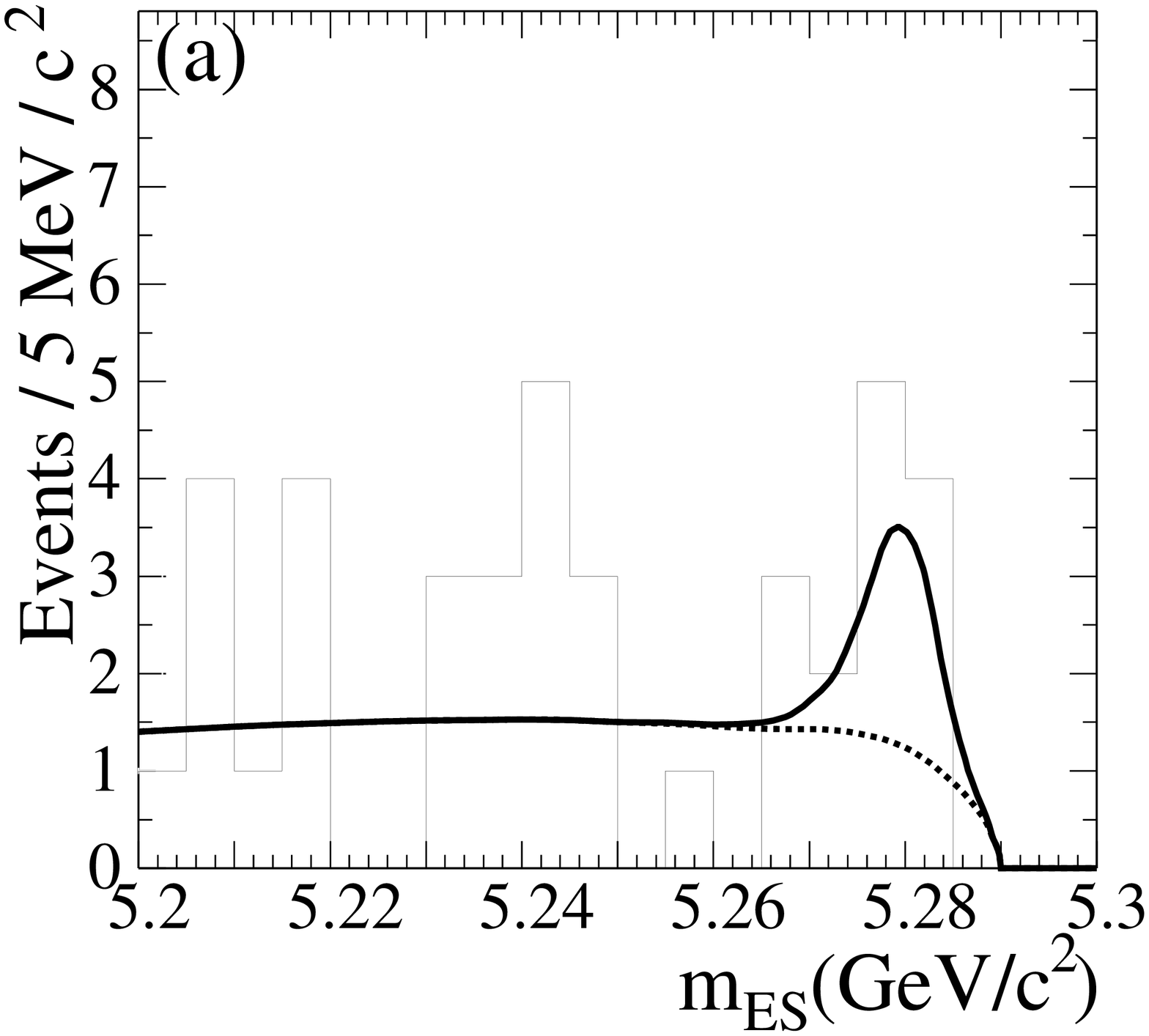}
\includegraphics[width=0.46\linewidth,height=0.483\linewidth]{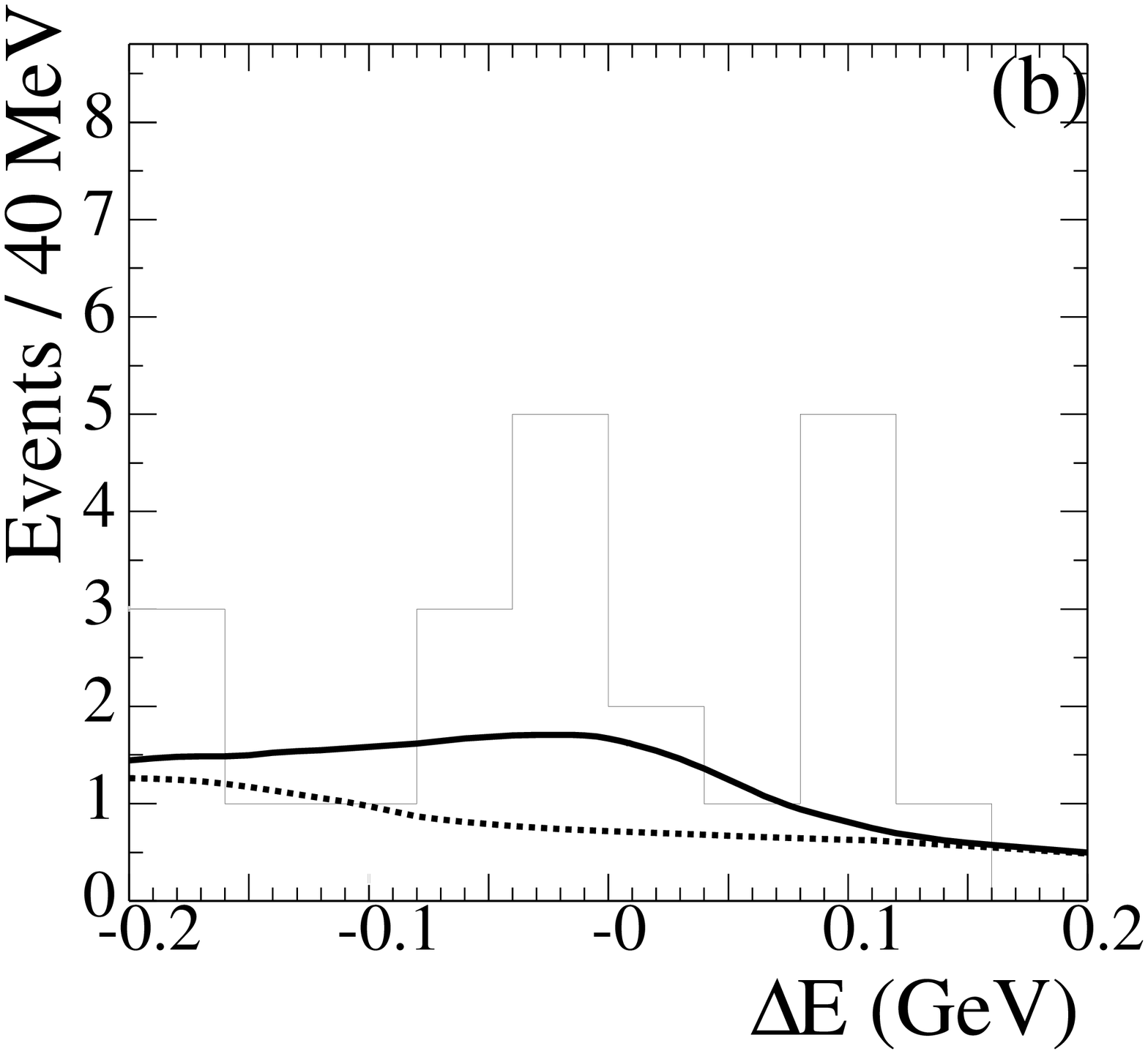}
\includegraphics[width=0.455\linewidth,height=0.483\linewidth]{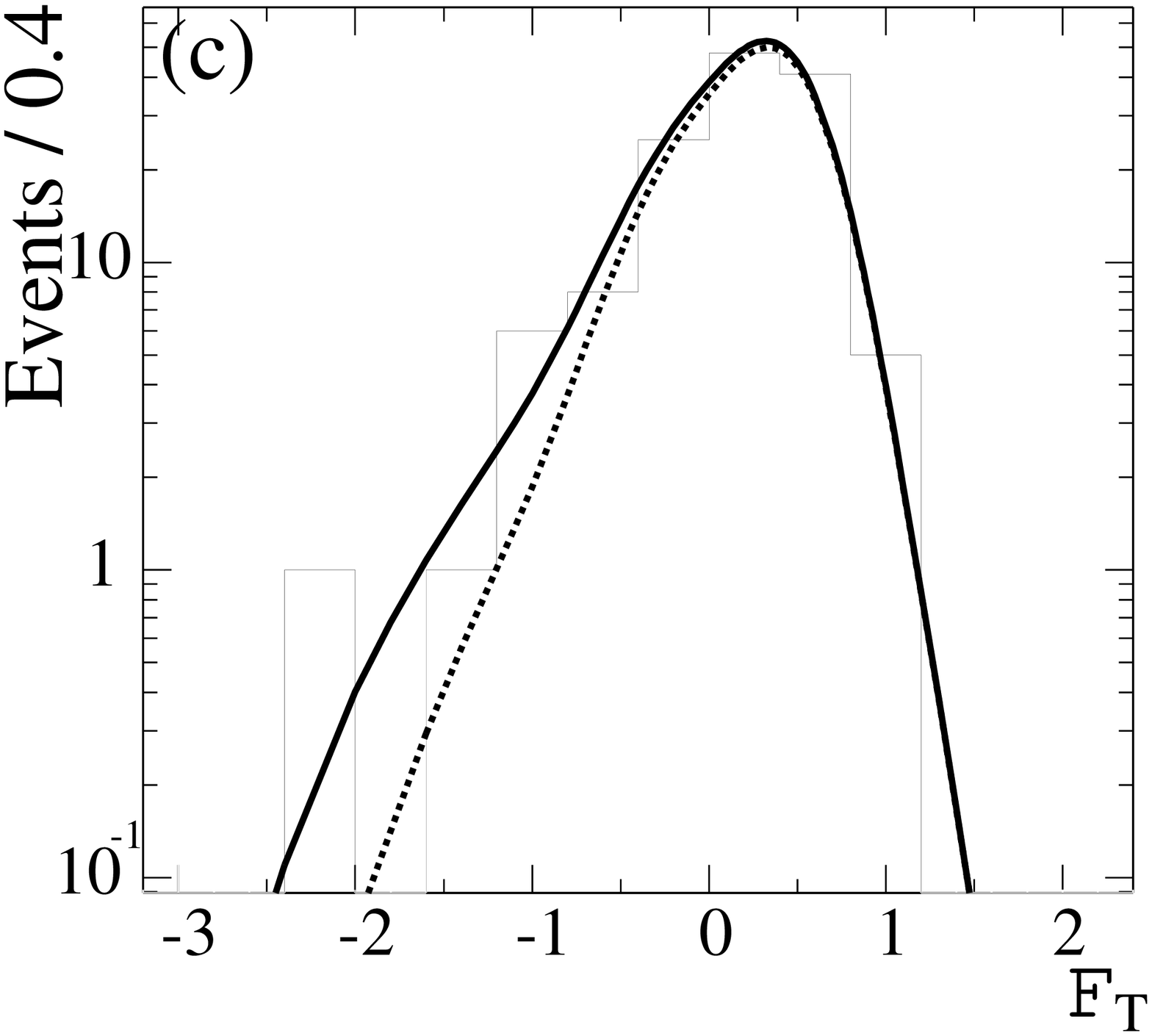}
\includegraphics[width=0.46\linewidth,height=0.482\linewidth]{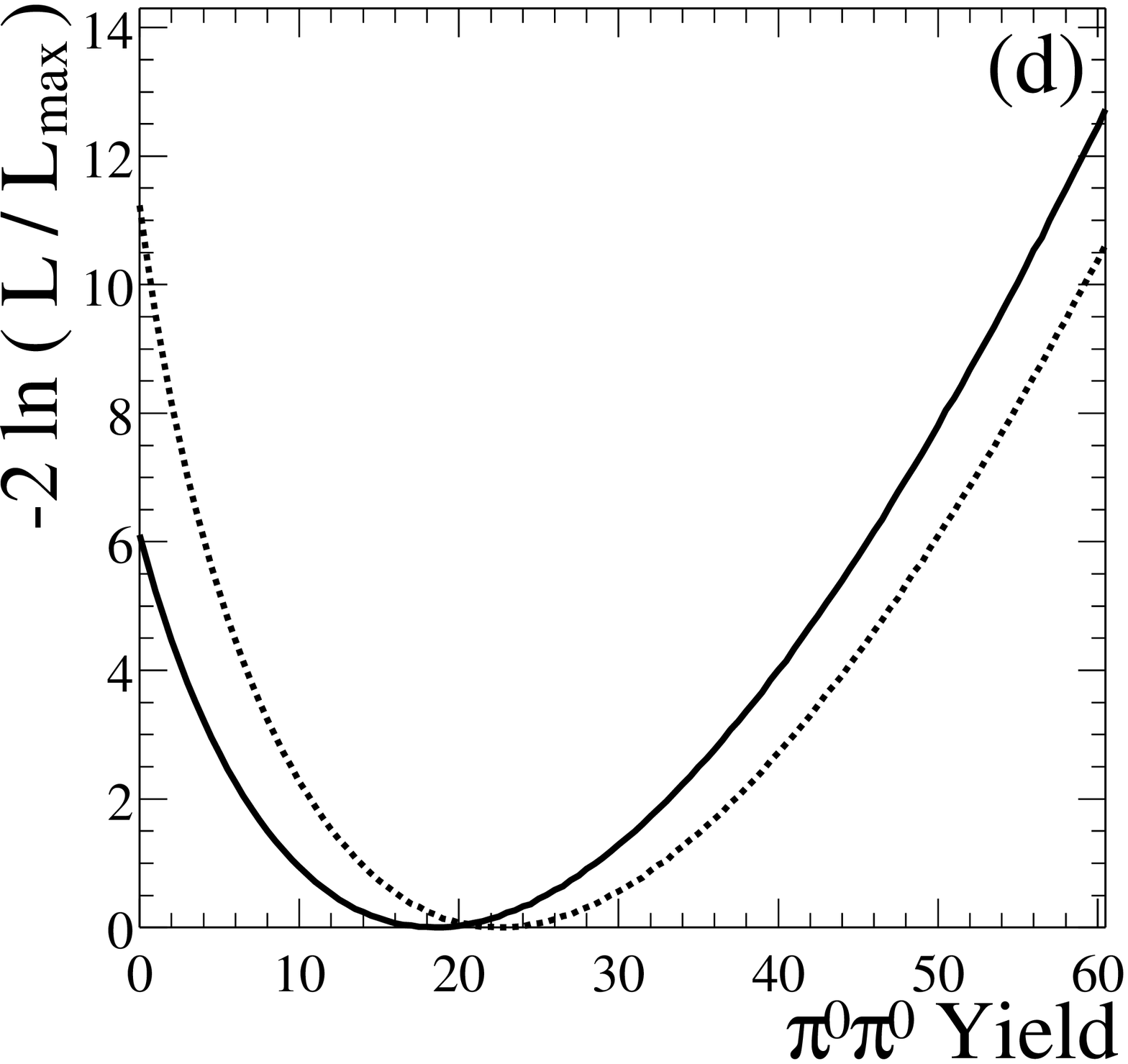}
\caption{The  a) \mes, b) \de, and c) \fishlt distributions for
\Bztopizpiz are shown, for  candidates
that satisfy optimized requirements on probability ratios
for signal to background  based on all variables except the one being
plotted.   The fraction of signal events included in the plots is
20\%, 20\% and 63\% for \mes, \de and \fishlt, respectively.
The dotted lines show the PDF projections for both \qqbar  and
\Btorhopiz background, while the solid lines are the PDF projections
for signal plus background.
The ratio
$-2\ln{(\mathcal{L}/\mathcal{L}_{max})}$ is shown in d) where the
dashed line is for statistical errors only and the solid line is
for statistical and systematic errors, as applied for the calculation
of significance.
}
\label{fig:pizpiz}
\end{center}
\end{figure}

The results of the maximum likelihood fits are summarized in
Table~\ref{table:summary}. Distributions of some of the variables used
in the fits are shown in Figs.~\ref{fig:hpiz} and \ref{fig:pizpiz} for
\Btohpiz and \Bztopizpiz, respectively.  The data shown are for events
that have passed a probability ratio cut optimized to enhance the
signal to background fraction.  The likelihood function for
\Bztopizpiz is shown in Fig.~\ref{fig:pizpiz}d.  The statistical
errors on the number of events are given by the change in signal yield
$n_i$ that corresponds to an increase in $-2\ln{\cal L}$ of one unit.
The systematic uncertainty in the likelihood fit is estimated by
varying the PDF parameters by their statistical errors or by comparing
the result with an alternate parameterization.  

For \Btopipiz, the
dominant systematic uncertainty is due to the \fish PDF for 
signal ($\pm 6.2$ events) and  background ($\pm 7.6$ events)
PDFs, while for \Btokpiz it is due to the \mes PDF for signal
($^{+2.7}_{-4.6}$ events).  Systematic uncertainties on the $\CP$
asymmetries are evaluated from PDF parameter variations and the upper limit on
intrinsic charge bias in the detector (1.0\%).

For \Bztopizpiz, systematic uncertainties from the PDFs are due to the 
\fishlt PDF for \qq background ($ ^{+7.5} _{-2.4}$ events), 
the \mes PDF for \qq background ($^{+1.2} _{-1.1}$ events), and the \de PDF for \qq background 
($^{+1.0} _{-0.2}$ events).   Additional systematic uncertainties for \Bztopizpiz arise from 
uncertainty in the EMC energy scale ($^{+0.8} _{-1.1}$ events), the \Btorhopiz rejection cut 
($\pm 1.3$ events), and uncertainty in the assumed \Btorhopiz branching fraction 
($^{+1.6} _{-1.9}$ events).  
 The significance of the event yield, also listed in Table~\ref{table:summary}, is evaluated
from the square root of the change in $-2\ln{\cal L}$ with the signal yield
fixed to zero. 
The upper limit for \Bztopizpiz is evaluated by finding $n_{\piz\piz}$ where
$ \int_{0}^{n_{\piz\piz}}\mathcal{L}(n) dn / \int_{0}^{\infty}\mathcal{L}(n) dn = 0.9$. For
both significance and upper limits, systematic uncertainties are included with a worst case 
assumption for efficiencies and PDF variations.

We observe  $\BR(\Btopipiz) = (5.5^{+1.0}_{-0.9} \pm 0.6)\times
10^{-6}$, with a statistical significance of  $7.7\sigma$ from zero.
This result is consistent with several prior measurements reporting
evidence for this
decay~\cite{babarcharmless,bellecharmless,cleocharmless}. We measure
$\BR(\Btokpiz) = (12.8^{+1.2}_{-1.1} \pm 1.0)\times 10^{-6}$. 
No evidence of direct \CP violation is observed.
Our limit $\BR(\Bztopizpiz) < 3.6 \times
10^{-6}$ improves upon prior results~\cite{cleopizpiz,bellecharmless}.
  Removing correlated
systematic uncertainties from luminosity and \piz efficiency, we bound
the ratio $\mathcal{B}(\Bztopizpiz)/\mathcal{B}(\Btopipiz) < 0.61$ at a 90\%
confidence level. Assuming isospin relations for $\Btopipi$~\cite{Isospin}, this corresponds to
an upper limit of $|\alpha_{\rm eff} - \alpha| < 51^{\rm o}$.

\par

We are grateful for the excellent luminosity and machine conditions
provided by our \pep2\ colleagues, 
and for the substantial dedicated effort from
the computing organizations that support \babar.
The collaborating institutions wish to thank 
SLAC for its support and kind hospitality. 
This work is supported by
DOE
and NSF (USA),
NSERC (Canada),
IHEP (China),
CEA and
CNRS-IN2P3
(France),
BMBF and DFG
(Germany),
INFN (Italy),
FOM (The Netherlands),
NFR (Norway),
MIST (Russia), and
PPARC (United Kingdom). 
Individuals have received support from the 
A.~P.~Sloan Foundation, 
Research Corporation,
and Alexander von Humboldt Foundation.


\begin{thebibliography}{99}
\bibitem{Argus} The threshold function used for the \mes PDF is $\frac{\mes}{\mz} \sqrt{1 -
(\frac{\mes}{\mz})^{2}} \, {\rm exp}\left\{-\xi\left[1 - (\frac{\mes}{\mz})^{2}\right]\right\}$,
where \mz is the \mes endpoint, and $\xi$ the shape parameter.  See 
 ARGUS Collaboration, H.\ Albrecht {\em et al.}, \zpc{48}, 543 (1990).
\end{thebibliography}
\end{document}